\documentclass[aps,pra,twocolumn,superscriptaddress]{revtex4-1}
\usepackage[latin9]{inputenc}
\usepackage{mathtools}
\usepackage{amsbsy}
\usepackage{amstext}
\usepackage[unicode=true,pdfusetitle,
 bookmarks=false,
 breaklinks=false,pdfborder={0 0 1},backref=false,colorlinks=false]
 {hyperref}
\usepackage[dvipsnames]{xcolor}
 \hypersetup{colorlinks=true,citecolor=NavyBlue,filecolor=black,linkcolor=black,urlcolor=black}

\makeatletter

\usepackage{amsmath}
\usepackage{graphicx}
\usepackage{amssymb}
\usepackage{txfonts,color}
\usepackage{dsfont}
\makeatother

\newcommand{\red}[1]{{\color{red}}}

\usepackage{pifont}

\newcommand{\ba}{\begin{eqnarray}}
\newcommand{\ea}{\end{eqnarray}}

\begin{document}

\title{Non-equilibrium readiness and accuracy of Gaussian Quantum Thermometers} 

\author{Luca Mancino}
\affiliation{Centre for Theoretical Atomic, Molecular and Optical Physics,
School of Mathematics and Physics, Queen's University, Belfast BT7 1NN, United Kingdom}

\author{Marco G. Genoni}
\affiliation{Quantum Technology Lab, Dipartimento di Fisica, Universit\`{a} degli Studi di Milano, 20133 Milano, Italy}

\author{Marco Barbieri}
\affiliation{Dipartimento di Scienze, Universit\`{a} degli Studi Roma Tre, Via della Vasca Navale 84, 00146, Rome, Italy}
\affiliation{Istituto Nazionale di Ottica, CNR-INO, 50125, Florence, Italy}

\author{Mauro Paternostro}
\affiliation{Centre for Theoretical Atomic, Molecular and Optical Physics,
School of Mathematics and Physics, Queen's University, Belfast BT7 1NN, United Kingdom}

\begin{abstract} 
The dimensionality of a thermometer is key in the design of quantum thermometry schemes. In general, the phenomenology that is typical of finite-dimensional quantum thermometry does not apply to infinite dimensional ones. We analyse the dynamical and  metrological features of non-equilibrium Gaussian Quantum Thermometers: on one hand, we highlight how quantum entanglement can enhance the readiness of composite Gaussian thermometers; on the other hand, we show that non-equilibrium conditions do not guarantee the best sensitivities in temperature estimation, thus suggesting the reassessment of the working principles of quantum thermometry.
\end{abstract}

\maketitle

\section{Introduction}
The direct assessment of the properties of quantum mechanical systems is not always possible or convenient: in general, any direct interference would alter the properties of the system, possibly spoiling them. This problem has motivated the research for effective schemes for {\it indirect quantum probing} that are able to provide information on the quantities of interest while only weakly affecting the system at hand \cite{Kelvin,JacobsBook}. Quantum thermometry, {\it i.e.} the estimate of the operating temperature of a quantum system, offers interesting opportunities for the design and application of indirect probing strategies, which would be useful for the 
characterisation and control of temperature of micro- and nano-devices ~\cite{Michalski2002,Giazotto2006,Brites2012}. 
\\
Most of the current investigations in quantum thermometry have used two-level systems as thermometers~\cite{Brunelli2011,Jevtic2015,Paris2015}, shedding light on the link between the equilibrium heat capacity of such microscopic probes and the amount of information that can be gathered on the temperature of the environment~\cite{Hill2001,DePasquale2016,Cavina2018}, introducing bounds on the irreversible entropy production of the probe~\cite{Deffner2011,Mancino2018,Mancino22018}, and clarifying the extent of the advantages resulting form finite-time interactions for both temperature discrimination and estimation~\cite{Cavina2018,Mancino2017}. 
Such investigations have ultimately opened the path to the exploration of the role played by genuine quantum features in the enhancement of the thermometric performance of two-level quantum probes ~\cite{Stace2010,Neumann2013,Kosloff13,Kucsko2013,Correa2015,Campbell2017,Campbell2018,Sbroscia2018,Seah2019,Farina2019,Jorgensen2020}. However, little has been explored about the effects that the dimensionality of the quantum probe has on the features of a given thermometric protocol.
\\
This is precisely the context within which the study reported in this article moves. We consider quantum thermometry operated using infinite-dimensional (Gaussian) quantum probes, each interacting with its own thermal bath, whose temperature we aim at estimating ~\cite{Monras2011,Pinel2013,Gao2014}. We show that entanglement between the probes significantly impacts on the readiness of the thermometer, thus marking significant discrepancies with the recently explored finite-dimensional case reported in Ref.~\cite{Feyles2019}, where geometric considerations on the dynamics of the thermometer have been explored. 

Here, we show that a composite thermometer operating under non-equilibrium conditions not always offers higher sensitivities in temperature estimation, thus leaving room for the reappraisal of some of the aspects underpinning quantum thermometry ~\cite{Cavina2018,Correa2015}. 
\\
While the formalism used to illustrate our findings is that of Gaussian quantum states and operations ~\cite{Ferraro2005,Serafini2017}, our study addresses a wealth of physical situations of strong experimental relevance for quantum probing, from micro-/nano-mechanical oscillators driven by optical or electric forces to microwave fields in superconducting waveguides and atomic-spin systems collectively coupled to driving fields ~\cite{Nichols2018}.

The remainder of this work is organised as follow: in Sec. II, we describe the model addressed in our analysis. Sec. III is devoted to geometric considerations leading to the definition of a dynamical speed on the Riemannian manifold of quantum states. In Sec. IV, we discuss the metrological accuracy of the probing mechanism through the evaluation of the non-equilibrium Quantum Fisher Information (QFI) of the system. Sec. V then summarises our findings and opens up new perspectives of investigation.

\section{The physical model}
We consider a thermometric scheme which is based on the weak interaction between two $N$-atom spin-systems (labelled as $a$ and $b$) and their respective thermal bath. Collective states of atomic spin systems have long been considered for metrological tasks in light of their high sensitivity~\cite{Hammerer}, and thus embody a natural platform where to investigate thermometry.
 
The ground and excited states of each atom are $\vert g \rangle$ and $\vert e \rangle$, respectively, and the free Hamiltonian of the probing system is $\hat{H}^0=\hbar \omega_a ( \sum_{k=1}^N \hat{\sigma}_{k,a}^z + 1/2 ) + \hbar \omega_b ( \sum_{k=1}^N \hat{\sigma}_{k,b}^z + 1/2)$, where $\omega_{a,b}$ represents the transition frequency of the atoms. The ensembles are coupled through the interaction term 
\begin{equation}
\hat{H}^{int}=\frac{\hbar G}{2} \hat S^x_a\hat S^x_b,
\label{Ham}
\end{equation}
 where we have introduced the collective spin operators $\hat{S}_j^x=\hat S^-_j + \hat S^+_j$ with $\hat{S}_j^\pm=\sum_{k=1}^N \hat{\sigma}_{k,j}^\pm$ 
 ($j=a,b$) and $\hat{\sigma}_k^+=\hat{\sigma}_k^{-\dag}=\vert e \rangle \langle g \vert_k$. Eq.~\eqref{Ham} results from the off-resonant dipole-like coupling of the two ensembles with the same light field, which is then adiabatically eliminated~\cite{Hammerer}.  As collective spin-systems are symmetric with respect to particle exchange, we can ignore any external degree of freedom associated with position of the individual particles.
 
As mentioned, local and independent baths -- coupled to the respective subsystem at a rate $k_j$ -- induce thermal fluctuations described introducing input operators $\hat{j}^{in}$, each characterised by the two-time correlation functions $\langle \hat{j}^{in,\dagger}(t) \hat{j}^{in}(t') \rangle=M_j \delta(t-t')$, and $\langle \hat{j}^{in}(t) \hat{j}^{in,\dagger}(t') \rangle=(M_j+1) \delta(t-t')$. Here, $M_j=(e^{\hbar \omega_a/k_B T_j}-1)^{-1}$ is the thermal occupation number of bath $j$ at temperature $T_{j}$ and $k_B$ is the Boltzmann constant \cite{Breuer2002}. 
 
We assume that the spin system exhibits no large fluctuations (enforced by the assumption of $M_j\lesssim1$), so that the collective spin operators can be mapped onto effective bosonic degree of freedom -- with creation and annihilation operators $\hat j$ and $\hat j^\dag~(j=a,b)$ respectively -- through a Holstein-Primakoff (HP) transform ~\cite{Holstein1940,Marcuzzi2016}. In this regime, the interaction Hamiltonian is recast into the form $\hat{H}_{hp}^{int}=\hbar GN/2 (\hat{a}^\dagger - \hat{a}^\dagger \hat{a}^\dagger \hat{a}/2N + h.c.)(\hat{b}^\dagger - \hat{b}^\dagger \hat{b}^\dagger \hat{b}/2N + h.c.)$: in this contribution, terms of order higher than quadratic are suppressed by at least a factor $1/N$.

\begin{figure}[t!]
\centering
\includegraphics[width=\columnwidth]{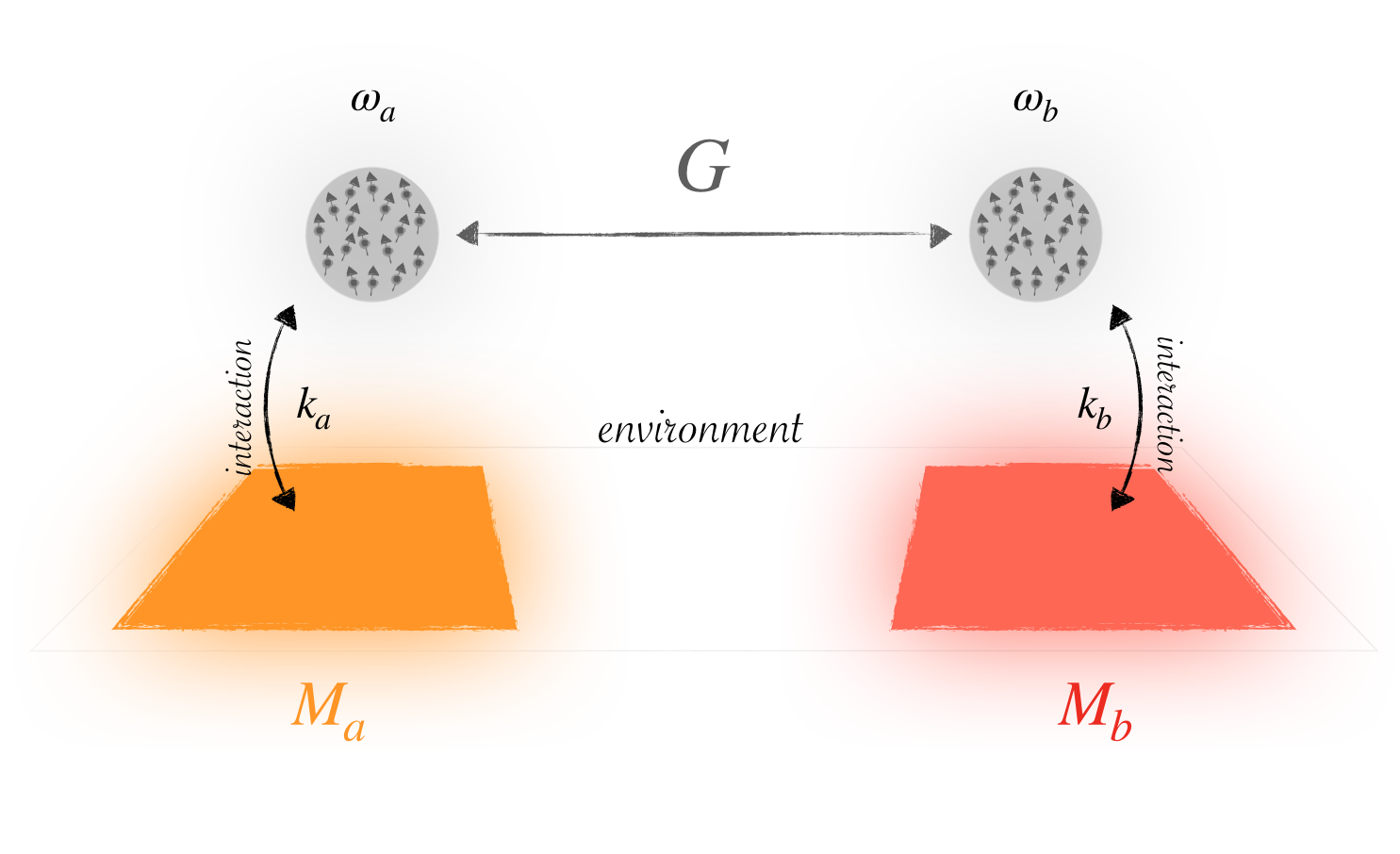}
  \caption{Conceptual scheme of the protocol. Two ensembles of atomic spin-states are led to interact with an environment which is made by two independent Markovian thermal baths at temperatures $T_a$ and $T_b$, with average numbers of excitations $M_a$ and $M_b$. The interaction between the two ensembles is modulated through the $G$ parameter. The dissipation to the local thermal bath depends on the $k_{a,b}$ decay rate.}
\label{Fig1}
\end{figure}

The dynamics is then described through the Heisenberg-Langevin equations 
\begin{equation}
\label{HL}
\partial_t \hat{u}_i (t) = \frac{i}{\hbar} [\hat{H}_{hp},\hat{u}_i (t)]+\hat{R}_{i}(t)
\end{equation}
with $H_{hp}$ the total Hamiltonian of the system in the HP regime, $\hat{{u}}=(\hat{a},\hat{a}^\dag,\hat{b},\hat{b}^\dagger)$ and $\hat{R}=(\hat{a}^{in},\hat{a}^{in,\dagger},\hat{b}^{in},\hat{b}^{in,\dagger})$. The presence of non-linear terms in $\hat{H}_{hp}$ makes the dynamics very difficult to tackle. 
In order to overcome this hurdle, we consider the fluctuations of the HP operators by taking the first-order expansion $\hat{j}=\bar{j}+\delta \hat{j}$, where $\bar{j} \in \mathbb{C}$ is a classical mean value and $\delta \hat{j}$ is a zero-mean fluctuation operator \cite{Paternostro2006}. The linearisation remains valid until the fluctuations of the HP operators are small compared to the steady-state values. On these grounds, we are legitimated to remove the non-linear contributions in the fluctuations from Eq.~\eqref{HL}. By introducing the vector of the dimensionless quantum fluctuations $\delta \hat{f} = (\delta \hat{x}_a, \delta \hat{p}_a, \delta \hat{x}_b, \delta \hat{p}_b)$, where $\delta \hat{x}_j = (\delta \hat{j}^\dagger + \delta \hat{j})/\sqrt{2}$, and $\delta \hat{p}_j = i (\delta \hat{j}^\dagger - \delta \hat{j})/\sqrt{2}$, we obtain the dynamical equations 
\begin{equation}
\begin{split}
& \partial_t \delta \hat{x}_a = \omega_a \delta \hat{p}_a - k_a \delta \hat{x}_a, \\
& \partial_t \delta \hat{p}_a = -\omega_a \delta \hat{x}_a - GN \delta \hat{x}_b - k_a \delta \hat{p}_a,  \\
& \partial_t \delta \hat{x}_b = \omega_b \delta \hat{p}_b - k_b \delta \hat{x}_b, \\
& \partial_t \delta \hat{p}_b = -\omega_b \delta \hat{x}_b - GN \delta \hat{x}_a - k_b \delta \hat{p}_b,
\label{heisenberglangevin}
\end{split}
\end{equation}
describing the dynamics of the quadrature operators for the two ensembles. Eq.~\eqref{heisenberglangevin} includes terms accounting for dissipation at rate $k_j$ into the local baths. 

\begin{figure*}[t!]
\centering
\includegraphics[width=1\textwidth]{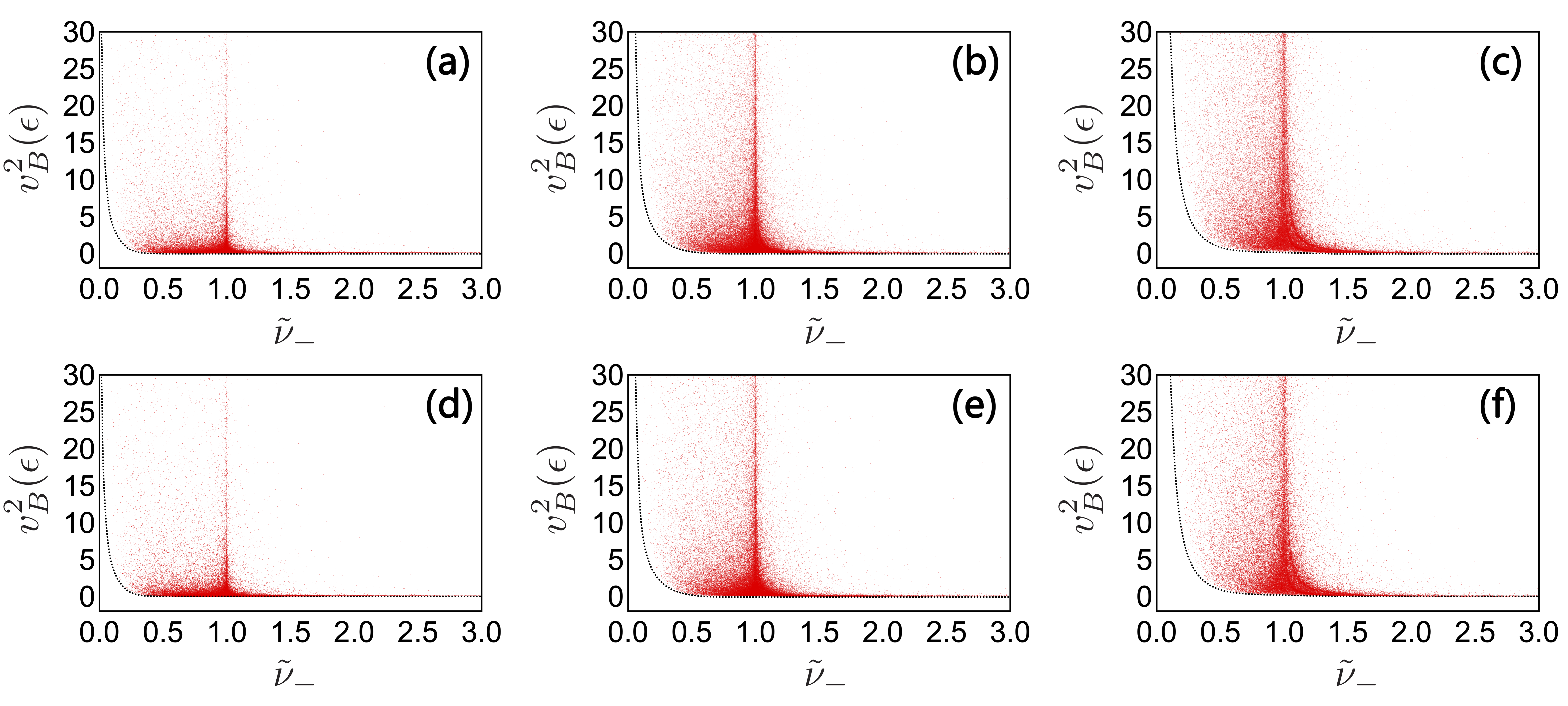}
  \caption{Behaviour of the Riemannian speed $v_{B}^2(\epsilon)$ \textit{vs} the symplectic eigenvalue $\tilde{\nu}_-$. The panels refer to different configurations of the system: in particular, $(a)-(c)$ $GN=0$, $(d)-(f)$ $GN=0.35$, $(a)(d)$ $M_a=M_b=0.1$, $(b)(e)$ $M_a=M_b=0.5$, $(c)(f)$ $M_a=M_b=1$. In the $\tilde{\nu}_- \leq 1$ entangled region, the lower bound on the Riemannian speed rears up more and more evidently when the temperature of the baths raises up.}
\label{Fig2}
\end{figure*}

The linearity of Eq.~\eqref{heisenberglangevin} and the choice of initial Gaussian states imply that the probability distribution describing the system is Gaussian at any time. Under these conditions, the complete description of the whole system can be enshrined in the covariance matrix of the fluctuations $\sigma_{mn}=\langle \{\delta \hat{f}_m, \delta \hat{f}_n\}\rangle$, as the first moments are always null. The equation of motion for the covariance matrix $\sigma$ takes the form of a deterministic diffusion Lyapunov equation $\partial_t \sigma = A \sigma + \sigma A^T + D$, where $A$ is the kernel matrix of the system's fluctuation operator, and $D=2k_a (2M_a + 1) \mathbb{I}_a \oplus 2k_b (2M_b + 1) \mathbb{I}_b$ is the diffusion matrix with $\mathbb{I}_j$ the $2 \times 2$ identity matrix. 

The stability of the solution is guaranteed by the Routh-Hurwitz test, ensuring that a unique non-equilibrium steady state described by the stationary covariance matrix $\sigma_{s}$ satisfying $A \sigma_{s} + \sigma_{s} A^T = - D$ is eventually achieved~\cite{Hurwitz1964}. If the whole system is non-interacting, each ensemble will equilibrate with its own thermal bath thus leading to a final thermal state. 

In what follows, we will exploit this model to explore the readiness of the composite Gaussian probe by introducing a geometric figure of merit able to define its dynamical speed. Moreover, we seek to investigate the metrological accuracy of such a non-equilibrium thermometer.

\section{Geometric considerations}
Here, we highlight the geometric features of the dynamics at hand making use of Riemannian tools. Specifically, we refer to the set of density matrices of a given quantum system, which form a Riemaniann manifold $\mathcal{S}$ over the Hilbert space $\mathcal{H}$ of the system: the one-to-one correspondence between the state of the system and its placement onto the manifold is such that, while evolving, the state will draw a curve $\ell_\gamma$ on $\mathcal{S}$, depending on $t$. For each $q \in \ell_\gamma$, an inner product on the tangent space $\mathbf{T}_q \mathcal{S}$ of the manifold $\mathcal{S}$ can be defined permitting the definition of an infinitesimal length $ds^2=Z_{\hat{\rho}}(d \hat{\rho},d \hat{\rho})$ \cite{Feyles2019,Bengtsson2006,Pires2016}. The reconstruction of an analytic form for the Riemannian metric $Z_{\hat{\rho}}$ in our problem is made difficult by the lack of a version of the Morozova-\v Cencov-Petz (MCP) theorem for infinite-dimensional systems~\cite{Morozova,Petz1996a,Petz1996b,Petz2002}. The MCP theorem states that each Riemannian metric for discrete variable systems is characterized by a correspondence with a set of functions that, by satisfying very restrictive conditions, can lead unambiguously to the quantum Fisher information metric ~\cite{uhlman1,uhlman2}. The restriction to Gaussian states, however, greatly simplifies the problem as the first and the second moments of the system are the only elements to be considered for its complete description \cite{Monras2010,Siudzinska2019}. 

Here, we focus on the Bures metric $ds_{B}^2=2[1-\mathcal{F}(\hat{\rho},\hat{\rho}+d\hat{\rho})]$, which can be computed through the Uhlmann fidelity $\mathcal{F}$ between two infinitesimally-close Gaussian states $\hat{\rho}$, and $\hat{\rho}+d\hat{\rho}$ \cite{Petz1996,Petz2002,Banchi2015}. By calling $r$ 
and $\sigma$ as the first and second moments of the state $\hat{\rho}$, the Bures metric can be expressed as $ds^2=(dr^T \sigma^{-1} dr)/2+\delta/8$, where $\delta=\rm{Tr}[d\sigma(\mathcal{L}_{\sigma} + \mathcal{L}_{\Omega})^{-1} d\sigma]$, with $\Omega$ as the $d$-mode symplectic matrix $\Omega=\oplus_{j=1}^d i\sigma_{y,j}$ (here, $d=2$ as we consider two-mode systems, and $\sigma_{y,j}$ is the $y$-Pauli matrix of the subsystem $j$), and $\mathcal{L}_Y X := YXY$ {for any pair of operators $X$ and $Y$}. The inverse operation on the superoperator refers to the Moore-Penrose pseudo-inverse. 

With simple algebra, the following form of the Bures metric can be obtained
\begin{equation}
ds_{B}^2=\frac{1}{4} \sum_{j=\pm} \frac{\left( \nu_j (t+dt) - \nu_j (t) \right)^2
}{\nu_j(t)^2-1},
\label{metricgaussian}
\end{equation}
where $\nu_\pm$ represent the symplectic eigenvalues of the Gaussian state. Eq.\eqref{metricgaussian} provides a simple and general way to manipulate the geometrical features of the considered Gaussian dynamics \cite{Banchi2015}. As a direct consequence of the Williamson's theorem, the covariance matrix of the subsystem $\sigma_{a,b}$ can be reduced in its standard form such that $\det \sigma_{a,b} = \nu_{a,b}^2$, with $\nu_{a,b}$ being the symplectic eigenvalue of the subsystem. In this way, the local form of the infinitesimal length $ds_{B,j}^2$ for the subsystem  $j=a,b$ can be obtained $
ds_{B,j}^2=\frac{1}{4} \frac{\left( \nu_j (t+dt) - \nu_j (t) \right)^2
}{\nu_j(t)^2-1}$, proving that $ds_{B}^2 \neq ds_{B,a}^2 + ds_{B,b}^2$. 

The above constructions lead to the instantaneous speed of quantum evolution on the Riemannian manifold
\begin{equation}
v_{B}^2(t)=\frac{1}{4} \sum_{j=\pm} \frac{\left( \partial_t \nu_j(t) \right)^2}{\nu_j(t)^2-1},
\label{speed}
\end{equation}
depending on the time derivatives of the symplectic eigenvalues $\nu_{\pm}$ of the Gaussian state. The form of the Riemannian speed in Eq.~\eqref{speed} can be applied whenever an active feedback mechanism suppressing the drift of the first moments is considered. It is noteworthy to stress that, under this condition, Eq.~\eqref{speed} does not depend on the specific dynamics at hand, and it is hence valid for general Gaussian closed and open system dynamics. 

\begin{figure*}[t!]
\centering
\includegraphics[width=1\textwidth]{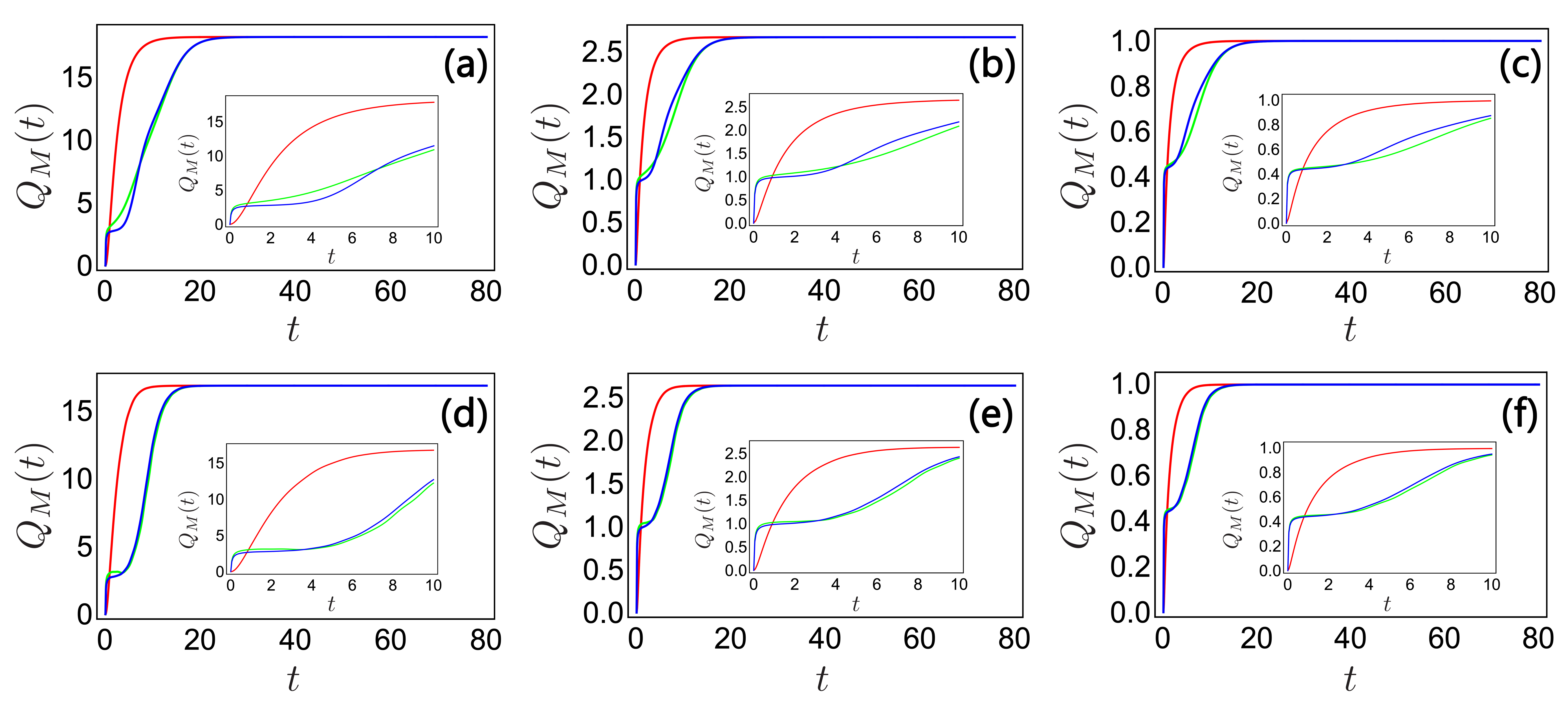}
  \caption{Plots of the non-equilibrium QFI $Q_M(t)$ for different initial preparations of the system as a function of the interaction time $t$. In all plots, red curves describe the behaviour of $Q_M(t)$ for initial thermal states $\sigma_0=(1+2\bar{M}_a)\mathbb{I}_2 \oplus (1+2\bar{M}_b)\mathbb{I}_2$ with $\bar{M}_a=\bar{M}_b=0.1$, blue curves show the same dispersion for locally squeezed vacuum states $\sigma_0=(S_a \sigma_{0,a} S_a^T) \oplus (S_b \sigma_{0,b} S_b^T)$ with $\sigma_{0,a}=\sigma_{0,b}=\mathbb{I}_2$ and $S_a$ ($S_b$) a squeezing transformation with parameter $r_a=2$ ($r_b=-2$), and green curves are associated to initial two-mode squeezed states (twin-beam states) with parameter $r=2$. The maximum of the QFI is reached when the system approaches its steady state, so that any absolute metrological advantage in the non-equilibrium regime is not observed. At the very beginning of the dynamics, the considered squeezed states show a neat advantage than their thermal counterpart: the rung characterising such an advantage is relatively more pronunced when the temperature of the bath is higher ((a)-(d) $M=0.1$, (b)-(e) $M=0.5$, and (c)-(f) $M=1$). The presence of the interaction between the two ensembles does not modify the general trend of the QFI ((a-c) $GN=0$, (d-f) $GN=0.35$).}
  \label{Fig3}
\end{figure*}

In what follows, we shed light on the consequences of non-classical correlations on the instantaneous Riemannian speed of the system. Being the latter a purely dynamical figure of merit, it is not explicitly connected to the accuracy of the thermometer at hand. On one side, such an inspection can be prospecting in unearthing dynamical aspects of quantum correlations, \textit{i.e.} entanglement, while on the other can lead to a more profound understanding of the dynamical readiness in the context of quantum thermometry. This exploration can be opportunely managed numerically, as correlation quantifiers are non-linear functions of the state: in this regard, we proceed by randomly generating {a large number} of Gaussian states to explore the role of non-classical features in the initial Riemannian speed of the system. 

The results of our numerical analysis are shown in Fig.~\ref{Fig2}, where we considered the symmetric case of $M_{a,b}=M$. There, we report the Riemannian speed of Gaussian states undergoing the dynamics shown in Eq.~\eqref{heisenberglangevin} for an arbitrarily small time $\epsilon$ versus the smallest symplectic eigenvalue $\tilde{\nu}_-$ of the covariance matrix associated with the partially transposed state of the system. According to the CV counterpart of the Peres-Horodecki criterion, entangled (separable) states have $\tilde{\nu}_- < 1$ ($\tilde{\nu}_- \ge 1$) states ~\cite{Giedke2001}. In contrast to what emerges from qubit-based thermometry \cite{Feyles2019}, the termometric phenomenology of CV entangled states is different from that resulting from the use of separable states. In particular, the trends displayed by Fig.~\ref{Fig2} suggest that larger degrees of entanglement -- i.e. smaller values of $\tilde{\nu}_-$ -- correspond to growing values of the minimum initial Riemannian speed. Statistically, entangled states appear to increase the initial readiness of the thermometer. Differently from qubit-based thermometry, the physical origin of this feature seems to be intrinsically related to non-classical effects. The numerical analysis points out the dependence of the lower bounds to the temperature of the external environment: in the entangled region, the larger the temperature of the baths, the larger the minimal initial Riemannian speed of the thermometer. Moreover, it is remarkable to highlight that  maximally entangled Gaussian states at fixed global and local purities approach the lower bound to the Riemannian speed at a given degree of entanglement, thus having the slowest response to the dynamics induced by the baths. Such states can be expressed as entangled two-mode squeezed thermal states, whose free parameters can be suitably minimised to numerically build the lower bound shown in Fig.~\ref{Fig2}. The presence of the interaction ($G \neq 0$) between the two subsystems does not affect the dynamical behaviour of the thermometer in Fig.~\ref{Fig2} as it is prospected to emerge at longer times.

\section{Metrological considerations}
As a second point of our investigation, we aim at exploiting the designed setup to build indirect measurement schemes able to infer temperature-related parameters associated to the environment. As we are probing a property of the bath, which is maintained in its initial thermal state, the temperature here is a well-defined parameter throughout the evolution. In other words, we are not concerned with an effective temperature of the probe, nor we need to be. In particular, the case at hand is ideal for this purpose as the adopted system can be initialised with a high degree of control, and then effectively measured after its interaction with the sample to be probed.
In the adopted scheme, we consider $\mathcal{N}$ independent copies of the system at hand, allowing for the same number of independent measurements which can be implemented to obtain an unbiased estimator $\tilde{M}$ for the average number of excitations in the baths $M$. In this case, the variance $\Delta^2 M=\langle (\tilde{M}-M)^2 \rangle$ satisfies the Quantum Cram\'{e}r-Rao bound (QCRB) $\Delta^2 M \geq 1/\mathcal{N} Q_M$, where $Q_M$ is the Quantum Fisher Information (QFI). Plainly, the accuracy of the measurement can be enhanced by increasing the QFI \cite{Giovannetti2004,Giovannetti2006}. 
The latter can be promptly interpreted as the \textit{distance} between Gaussian states interacting with thermal baths whose average numbers of excitations differ by an infinitesimal variation, so that $Q_M(t)=8(1-\mathcal{F}(\hat{\rho}_M(t), \hat{\rho}_{M+dM}(t)))/dM^2$, where the functional $\mathcal{F}$ is the Ulhmann fidelity between the Gaussian states of the thermometer $\hat{\rho}_M$ and $\hat{\rho}_{M+dM}$ which are respectively interacting with environments at temperature $M$ and $M+dM$. Remarkably for Gaussian states, the evaluation of the QFI can be traced back to the Wigner formalism being first and second moments sufficient for the complete description of the state \cite{Banchi2015}. In the specific case at hand, where first moments can be flashed out from the computation, the fidelity can be defined as 
\begin{equation}
\mathcal{F}^4(\hat{\rho}_M(t), \hat{\rho}_{M+dM}(t)) = \frac{ \text{det} \left[ 2 \left( \sqrt{\mathbb{I}+\frac{(\Xi(t) \Omega)^{-2}}{4}} + \mathbb{I} \right) \Xi(t) \right] }{ \text{det} \left[ \frac{\sigma_M(t)+\sigma_{M+dM}(t)}{2} \right] },
\end{equation}
where $2\Omega \Xi(t)=(\sigma_M(t) + \sigma_{M+dM}(t))^{-1} (\Omega + \sigma_{M+dM}(t) \Omega \sigma_M(t))$. 

A reparametrisation of the QFI is made possible by the fact that $M$ is a continuously differentiable function of $T$. In this spirit, the QFI on the temperature can be expressed as 
\begin{equation}
Q_T(t)=Q_M(t) \frac{\omega^2 \text{csch}^4 \left( \frac{\omega}{2T} \right)}{16 T^4}.
\end{equation}
The plots in Fig.~\ref{Fig3} show the dispersion of the non-equilibrium QFI $Q_M(t)$ as a function of the interaction time $t$. At variance with the findings in Ref.\cite{Jevtic2015,Cavina2018,Mancino2017} where qubit-based thermometers have been considered, we notice how non-equilibrium conditions do not enhance the accuracy of temperature estimation in the Gaussian regime. In particular, $Q_M(t)$ reaches its maximum value when the thermometer approaches its steady state. Fig.~\ref{Fig3} shows that thermometers prepared in a thermal state allow to get the maximum of the QFI in less time, if compared to initial locally squeezed or two-mode squeezed states. The presence of non-classical resources in the initial state of the thermometer brakes the achievement of the maximum in the QFI, but powers up its accuracy for very short times. Entanglement does not appear to have a clear role in enhancing the performance of the probe: indeed, thermometers prepared in two-mode squeezed states can lead to both increased and decreased metrological performances if compared to those prepared in locally squeezed vacuum states. In particular, the quantum-enhancement in temperature estimation due to the presence of the non-classical correlation is well displayed at short times, where entanglement leads up the QFI. It is remarkable to highlight that such an advantage appears more pronounced for low temperatures of the external thermal bath, and deteriorates the accuracy of the thermometer in the remaining transient of the non-equilibrium dynamics. The presence of the interaction between the subsystems accelerates the attainment of the optimal metrological condition, and reduces the gap between two-mode squeezed states, and locally squeezed vacuum states.

\section{Conclusions}
We have shown that the link between quantum features and facets of quantum thermometry are often elusive. By addressing explicitly the CV regime, we have highlighted a profound discrepancy in the way small-scale thermometers are influenced by quantum features. In the examined case, the instantaneous response of the Gaussian thermometer to the interaction with the environment shows a dependence on non-classical correlations among probing subsystems: the minimum initial dynamical speed of the system raises up with the amount of entanglement and depends on the temperature of the bath. No exceptions for the performance of the thermometer, which can be powered up for very short times by using quantum resources. The latter, however, can be detrimental for the remainder transient dynamics of the thermometer, reducing the non-equilibrium QFI with respect to classical states. The  picture coming out of our assessment pushes for further exploration of the role of quantum facets in quantum thermometry. The presence of higher-order non-linearities in the interaction Hamiltonian of the subsystems ought to be taken in consideration to investigate unexplored thermometric regimes. 

\bigskip

{\it Note}: During preparation of this manuscript, we became aware that similar considerations have been independently highlighted by I. Gianani et al. 

\section*{Acknowledgments}
The authors would like to thank Antonella De Pasquale, Gabriele De Chiara, Gabriel Landi, Obinna Abah, Michele Maria Feyles, and Federico Centrone for stimulating discussions. LM acknowledges the hospitality of Universit\`{a} degli Studi Roma Tre, where part of this work has been carried out. The authors acknowledge financial support from H2020 through the Collaborative Project TEQ (Grant Agreement No. 766900), and the Angelo Della Riccia Foundation (R.D. 19.7.41. n.979, Florence).

\bigskip

\section*{Appendix}

\subsection*{Holstein-Primakoff transformation}
In this Appendix, we briefly show how to manipulate the Hamiltonian of the whole system through the Holstein-Primakoff transformation. 

Assuming that $N \gg 1$, the Holstein-Primakoff transformation to boson creation and annihilation operators is defined by these relations
\begin{equation}
\hat{S}_j^+=\sqrt{N} \hat{j}^\dagger \hat{J}, \;\; \hat{S}_j^-=\sqrt{N} \hat{J} \hat{j}, \;\; \hat{S}_j^z= \hat{j}^\dagger \hat{j}-\frac{N}{2},
\label{dickeoperators}
\end{equation}
where $[\hat{j},\hat{j^\dagger}]=1$, and $\hat{J}=\sqrt{1-\hat{j}^\dagger \hat{j}/N}$ ($\hat{A}$ and $\hat{B}$ for the considered ensembles) is an operator ensuring that the operators in Eq.~\eqref{dickeoperators} fulfil the SU(2) algebra. It is fair to say that the operators in Eq.\eqref{dickeoperators} can be referred to the specific ensemble of atomic spin-states by substituting $\hat{a}$ and $\hat{b}$ to the generic operator $\hat{j}$. In terms of the collective bosonic Dicke operators, the Hamiltonian of the system $\hat{H}_{hp}$ results to be the sum of the free Hamiltonian $\hat{H}_{hp}^{0}=\hbar \omega_a (\hat{a}^\dagger \hat{a}-(N-1)/2) + \hbar \omega_b (\hat{b}^\dagger \hat{b}-(N-1)/2)$, and the interaction one $\hat{H}_{hp}^{int}=\hbar G/2 (\sqrt{N} \hat{a}^\dagger \hat{A} + \sqrt{N} \hat{A} \hat{a})(\sqrt{N} \hat{b}^\dagger \hat{B} + \sqrt{N} \hat{B} \hat{b})$. The presence or not of non-linear effects from the $\hat{J}$-like operators results from a trade-off between $\langle \hat{j}^\dagger \hat{j} \rangle$ and $N$. In general, the operators in Eq.\eqref{dickeoperators} can be manipulated through the $\hat{J}=\sum_{h=0}^\infty \frac{(2h)!}{(1-2h)(2^h h!)} \frac{(\hat{j}^\dagger \hat{j})^h}{N^h}$ expansion: here, we limit the expansion to $\hat{J} \sim 1 - \hat{j}^\dagger \hat{j}/2N + \mathcal{O}(N^{-2})$ which leads to the non-linear interaction Hamiltonian $\hat{H}_{hp}^{int}=\hbar GN/2 (\hat{a}^\dagger - \hat{a}^\dagger \hat{a}^\dagger \hat{a}/2N + h.c.)(\hat{b}^\dagger - \hat{b}^\dagger \hat{b}^\dagger \hat{b}/2N + h.c.)$, corresponding to the Hamiltonian used in the main text. It is remarkable to verify that the structure of this interaction Hamiltonian reduces to the one describing a position-like interaction between harmonic oscillators when dropping the non-linear contributions from the expansion. 

In order to account for the open dynamics of the thermometer, we make use of the Heisenberg-Langevin equations $\partial_t \hat{\mathcal{O}}=\frac{i}{\hbar}[\hat{H}_{hp},\hat{\mathcal{O}}]+\hat{N}$, with $\hat{\mathcal{O}}$ as a generic operator of the system. The problem of determining the dynamics of the probe reduces to the evaluation of the commutators $[\hat{H}_{hp},\hat{\mathcal{O}}]$. The contributions to the dynamics arising from the evalutation of these commutators can be highly non-linear: a way to overcome this roadblock consists in referring to the fluctuations $\delta \hat{\mathcal{O}}$ of the $\hat{\mathcal{O}}$ operators defined through the $\hat{\mathcal{O}}=\mathcal{O}_s + \delta \hat{\mathcal{O}}$ expansion. In this way, we are legitimate to drop out the non-linear contributions in the fluctuation operators, and the problem greatly simplifies. In what follows, we report the computation of the $[\hat{H}_{hp},\delta \hat{x}_j]$ commutator resulting in 
\begin{equation}
[\hat{H}_{hp},\delta \hat{x}_a]=- i \hbar \omega \delta \hat{p}_a, \;\; [\hat{H}_{hp},\delta \hat{x}_b]=- i \hbar \omega \delta \hat{p}_b.
\end{equation}
The commutators of the $[\hat{H}_{hp},\delta \hat{p}_j]$ quadrature position fluctuation operators have the following form
\begin{equation}
[\hat{H}_{hp},\delta \hat{p}_a]=i \hbar GN \hbar \delta \hat{x}_b + i \hbar \omega_a \delta \hat{x}_a, \;\; \text{and}
\end{equation}
\begin{equation}
[\hat{H}_{hp},\delta \hat{p}_b]=i \hbar GN \hbar \delta \hat{x}_a + i \hbar \omega_b \delta \hat{x}_b.
\end{equation}
The contribution of the \textit{classical} terms to the Heisenberg-Langevin equations is negligible when considering $\alpha_s(0)=\beta_s(0)=0$. 

\subsection*{Geometrical considerations: Bures metric}
The Bures metric between two infinitesimally close Gaussian states can be defined as $ds_B^2=2[1-\mathcal{F}(\hat{\rho},\hat{\rho}+d\hat{\rho})]$. Exploiting the Gaussian peculiarities of the states at hand, and considering a feedback mechanism suppressing the drift of the first moments in the phase-space, the Bures metric reduced to $ds_B^2=\delta/8$. In the case of a two-mode Gaussian state, a simple way to evaluate the Bures metric consists in building the Simon invariant form of the covariance matrix
\begin{equation}
\sigma=\begin{pmatrix}
\alpha & \gamma \\
\gamma^T & \beta
\end{pmatrix},
\end{equation}
where $\alpha=\text{diag}[a,a]$, $\beta=\text{diag}[b,b]$, and $\gamma=\text{diag}[c_+,c_-]$. Moving to the basis diagonalising $W = - i \sigma \Omega$, the form of $\delta$ greatly simplifies \cite{Banchi2015}
\begin{equation}
\delta=\sum_{ij} \frac{dw_i dw_j}{w_i w_j - 1},
\label{deltaBures}
\end{equation}
where $w_i$-s represent the eigenvalues of $W$, and $dw_i$-s represent their deviations. By exploiting the Binet theorem for which $\det W = \det (-i\Omega) \det \sigma = (-i)^{2 d} \det \sigma$ ($d$ being the number of modes), the spectrum of $W$ can be expressed in terms of the symplectic eigenvalues $\nu_\pm$ of $\sigma$. In the case at hand, $w_i$ is an element of $\text{Spec}[W]=(-i\sqrt{-\nu_+^2},+i\sqrt{-\nu_+^2},-i\sqrt{-\nu_-^2},+i\sqrt{-\nu_-^2})$, where 
\begin{equation}
\nu_\pm^2=\frac{1}{2} \left[ \Delta \pm \sqrt{\Delta^2 - 4 \det \sigma} \right],
\end{equation}
with $\Delta=\det \alpha + \det \beta + 2\det \gamma$. The expression of $W$ can be thus recast as $W=\text{diag}[-i\sqrt{-\nu_+^2 i^2},+i\sqrt{-\nu_+^2 i^2},-i\sqrt{-\nu_-^2 i^2},+i\sqrt{-\nu_+^2 i^2}]$, such that the differential of one of its components will correspond to $dw_i=-i\sqrt{-\nu_+^2(t+dt) i^2} + i\sqrt{-\nu_+^2 (t) i^2}$. Thus, the Bures metric can be expressed as
\begin{equation}
ds_B^2 = \sum_{ij} \frac{\delta_{ij}}{8} \frac{1}{w_i w_j - 1} dw_i dw_j,
\end{equation}
where $dw_i=w_i(t+dt) - w_i(t)$ (same expression for $dw_j$). 

By deforming the simplex, thus introducing the new coordinate
\begin{equation}
\Gamma_i= \frac{1}{4\sqrt2} \log \left( \frac{ \sqrt{w_i w_i -1}+w_i}{\sqrt{w_i w_i -1}-w_i} \right), 
\end{equation}
such that $d\Gamma_i=(d\Gamma_\alpha/d w_\alpha) dw_i$, with $d\Gamma_\alpha/d w_\alpha=(2\sqrt{2} \sqrt{w_\alpha w_\alpha - 1})^{-1}$, it is possible to gain a more compact version of the Bures metric $ds_{B}^2=\sum_i {d\Gamma_i} {d\Gamma_i}$, with which the Riemannian speed results to be $v_B^2(t)=\sum_i \frac{d\Gamma_i}{dt} \frac{d\Gamma_i}{dt}$.

Similar considerations can be done for the evaluation of the local Riemannian speed. In this case, the metric can be written as $ds_{B,j}^2=\frac{1}{4} \frac{(\nu_j(t+dt)-\nu_j(t))^2}{\nu^2(t)-1}$ so that $ds_B^2 \neq ds_{B,a}^2 + ds_{B,b}^2$.

\subsection*{Generation of random Gaussian states}
In order to randomly generate Gaussian states, we reduced the initial covariance matrices in the $Sp_{2,\mathbb{R}} \oplus Sp_{2,\mathbb{R}}$ Simon invariant form. Accordingly to \cite{Serafini2017}, the positivity of the covariance matrix $\sigma$, as well as the Robertson-Schr\"{o}dinger uncertainty relation $(\sigma + i \Omega) \geq 0$ (reducing to $\Delta \leq 1 + \det \sigma$ for two-mode states) dictate the physicality of the state. The $Sp_{4,\mathbb{R}}$ covariances can be parametrised through the marginal purities of the state $\mu_1=(\det \alpha)^{-1/2}=1/a$, $\mu_2=(\det \alpha)^{-1/2}=1/b$, and in terms of $\mu=(\det \sigma)^{-1/2}=[(ab-c_+^2)(ab-c_-^2)]^{-1/2}$, and $\Delta=a^2+b+2+2c_+ c_-$ which are, respectively, the global purity and the seralian of the state. With simple algebra, the following expression $c_\pm=\frac{\sqrt{\mu_1 \mu_2}}{4} (\eta_- \mp \eta_+)$ emerges, with $\eta_\mp=\sqrt{[ \Delta(\sigma) - {(\mu_1 \mp \mu_2)^2}/{\mu_1^2 \mu_2^2} ]^2 - {4}/{\mu^2}}$. From both the Robertson-Schr\"odinger condition and the presence of the $\eta_\pm$ radicals, the following conditions on the $Sp_{4,\mathbb{R}}$ symplectic invariants $0 \leq \mu_1 \leq 1$, $0 \leq \mu_2 \leq 1$, $\mu_1 \mu_2 \leq \mu \leq \frac{\mu_1 \mu_2}{\mu_1 \mu_2 + \vert \mu_1 - \mu_2 \vert}$, and $2/\mu + (\mu_1 - \mu_2)^2/\mu_1^2 \mu_2^2 \leq \Delta \leq \min \lbrace (\mu_1 + \mu_2)^2/ \mu_1^2 \mu_2^2 - (2/\mu), 1+ (1/\mu^2) \rbrace$ hold. In order to determine if the Gaussian state is entangled or not, the PPT condition can be applied on $\sigma$ thus leading to $\tilde{\sigma}$ where $\gamma \rightarrow \tilde{\gamma}$, with $\det \tilde{\gamma} = - \det \gamma$, and $\Delta \rightarrow \tilde{\Delta}=\det \alpha + \det \beta + 2 \det \tilde{\gamma}$ so that $\tilde{\Delta}=\Delta - 4 \det \gamma$. The symplectic eigenvalues of the partial transposed $\tilde{\sigma}$ covariance matrix can be expressed as $2\nu_\mp^2=\tilde{\Delta} \mp \sqrt{\tilde{\Delta}^2 - 4/\mu^2}$. From these conditions, the PPT criterion can be reduced to $\tilde{\Delta} \leq 1 + 1/\mu^2$ delimiting the region of separable (entangled) states, \textit{i.e.} those for which $\tilde{\nu}_- \geq 1 \; (\tilde{\nu}_- < 1)$. Such conditions allow for the classifications of Gaussian states in terms of their separabilty properties. In particular, the conditions $\mu < \mu_1 \mu_2$ as well as $\mu > \mu_1 \mu_2 /(\mu_1 \mu_2 + \vert \mu_1 - \mu_2 \vert)$ identify unphysical regions for Gaussian states, while $\mu_1 \mu_2 \leq \mu \leq \mu_1 \mu_2 / (\mu_1 + \mu_2 - \mu_1 \mu_2)$ is the region of Gaussian separable states. The remaining values for $\mu$ can be opportunely splitted to mark a coexistence region for $\mu_1 \mu_2 / (\mu_1 + \mu_2 - \mu_1 \mu_2) < \mu \leq \mu_1 \mu_2 / \sqrt{\mu_1^2 + \mu_2^2 - \mu_1^2 \mu_2^2}$, and an entangled one for the remaining physical values of the global purity. 

When fixing both global and local purities, Gaussian maximally entangled states can be obtained saturating the lower bound of the Seralian range of values $\Delta=2/\mu + (\mu_1 - \mu_2)^2/\mu_1^2 \mu_2^2$. The Gaussian maximally entangled states for fixed global and local purities (GMEMS) can be parametrised through $c_\pm=\pm (1/\mu_1 \mu_2 - 1/\mu)^{1/2}$. In the space of the entangled states, GMEMS have shown to be the slowest ones as they approach the lower bound in the Riemannian speed appearing in Fig.\ref{Fig2}.


\begin{thebibliography}{99}
\bibitem{Kelvin} Lord Kelvin famously stated that "{\it when you can measure what you are speaking about, and express it in numbers, you know something about it; but when you cannot measure it, when you cannot express it in numbers, your knowledge is of a meagre and unsatisfactory kind: it may be the beginning of knowledge, but you have scarcely, in your thoughts, advanced to the stage of science}" (Thomson 1889:73).
\bibitem{JacobsBook} K. Jacobs, {\it Quantum Measurement Theory and its Applications}, Cambridge University Press (2014).
\bibitem{Michalski2002} L. Michalski, K. Eckersdorf, J. Kucharski, and J. McGhee, {\it Temperature Measurement}, Measurement Science and Technology, 13, 10 (2002).
\bibitem{Giazotto2006} F. Giazotto, T.T. Heikkl\"{a}, A. Luukanen, A. M. Savin, and J. Pekola, {\it Opportunities for mesoscopics in thermometry and refrigeration: Physics and applications}, Rev. Mod. Phys. 78, 217 (2006).
\bibitem{Brites2012} C.D.S. Brites, P.P. Lima, N.J.O. Silva, A. Mill\'{a}n, V.S. Amaral, F. Palacio, and L.D. Carlos, {\it Thermometry at the nanoscale}, Nanoscale, 4, 4799-4829 (2012).
\bibitem{Brunelli2011} M. Brunelli, S. Olivares, and M.G.A. Paris, {\it Qubit thermometry for micromechanical resonators}, Phys. Rev. A 84, 032105 (2011).
\bibitem{Jevtic2015} S. Jevtic, D. Newman, T. Rudolph, and T.M. Stace, {\it Single-qubit thermometry}, Phys. Rev. A 91, 012331 (2015).
\bibitem{Paris2015} M.G.A. Paris, {\it Achieving the Landau bound to precision of quantum thermometry in systems with vanishing gap}, Jour. Phys. A: Math. and Theor., 49, 3 (2015).
\bibitem{Seveso2018} L. Seveso, and M.G.A. Paris, {\it Trade-off between information and disturbance in qubit thermometry}, Phys. Rev. A 97, 032129 (2018).
\bibitem{Hill2001} T.L. Hill, {\it A different approach to nanothermodynamics}, Nano Lett. 1, 273-275 (2001).
\bibitem{DePasquale2016} A. De Pasquale, D. Rossini, R. Fazio, and V. Giovannetti, {\it Local quantum thermal susceptibility}, Nat. Commun. 7, 12782 (2016).
\bibitem{Cavina2018} V. Cavina, L. Mancino, A. De Pasquale, I. Gianani, M. Sbroscia, R.I. Booth, E. Roccia, R. Raimondi, V. Giovannetti, and M. Barbieri, {\it Bridging thermodynamics and metrology in non-equilibrium quantum thermometry}, Phys. Rev. A 98, 050101(R) (2018).
\bibitem{Deffner2011} S. Deffner, and E. Lutz, {\it Nonequilibrium Entropy Production for Open Quantum Systems}, Phys. Rev. Lett. 107, 140404 (2011). 
\bibitem{Mancino2018} L. Mancino, V. Cavina, A. De Pasquale, M. Sbroscia, R.I. Booth, E. Roccia, I. Gianani, V. Giovannetti, and M. Barbieri, {\it Geometrical bounds on Irreversibility in Open Quantum Systems}, Phys. Rev. Lett. 121, 160602 (2018).
\bibitem{Mancino22018} L. Mancino, M. Sbroscia, E. Roccia, I. Gianani, F. Somma, P. Mataloni, M. Paternostro, and M. Barbieri, {\it The entropic cost of quantum generalised measurements}, npj Quant. Information 4, 20 (2018).
\bibitem{Mancino2017} L. Mancino, M. Sbroscia, I. Gianani, E. Roccia, and M. Barbieri, {\it Quantum Simulation of Single-Qubit Thermometry Using Linear Optics}, Phys. Rev. Lett. 118, 1130502 (2017). 
\bibitem{Stace2010} T.M. Stace, {\it Quantum limits of thermometry}, Phys. Rev. A 82, 011611(R) (2010).
\bibitem{Neumann2013} P. Neumann, I. Jakobi, F. Dolde, C. Burk, R. Reuter, G. Waldherr, J. Honert, T. Wolf, A. Brunner, J.H. Shim, D. Suter, H. Sumiya, J. Isoya, and J. Wrachtrup, {\it High-Precision Nanoscale Temperature Sensing Using Single Defects in Diamond}, Nano Lett. 13, 6, 2738-2742 (2013).
\bibitem{Kosloff13} R. Kosloff, {\it Quantum Thermodynamics: A Dynamical Viewpoint}, Entropy, 15(6), 2100-2128 (2013);
\bibitem{Kucsko2013} G. Kucsko, P.C. Maurer, N.Y. Yao, M. Kubo, H.J. Noh, P.K. Lo, H. Park, and M.D. Lukin, {\it Nanometre-scale thermometry in a living cell}, Nature 500, 54-58 (2013).
\bibitem{Correa2015} L.A. Correa, M. Mehboudi, G. Adesso, and A. Sanpera, {\it Individual Quantum Probes for Optimal Thermometry}, Phys. Rev. Lett. 114, 220405 (2015).
\bibitem{Campbell2017} S. Campbell, M. Mehboudi, G. De Chiara, and M. Paternostro, {\it Global and local thermometry schemes in coupled quantum systems}, New. J. Phys. 19, 103003 (2017).
\bibitem{Campbell2018} S. Campbell, M.G. Genoni, and S. Deffner, {\it Precision thermometry and the quantum speed limit}, Quantum Sci. Technol. 3, 025002 (2018). 
\bibitem{Sbroscia2018} M. Sbroscia, I. Gianani, L. Mancino, E. Roccia, Z. Huang, L. Maccone, C. Macchiavello, and M. Barbieri, {\it Experimental ancilla-assisted phase estimation in a noisy channel}, Phys. Rev. A 97, 032305 (2018).
\bibitem{Seah2019} S. Seah, S. Nimmrichter, D. Grimmer, J.P. Santos, V. Scarani, and G.T. Landi, {\it Collisional Quantum Thermometry}, Phys. Rev. Lett. 123, 180602 (2019).
\bibitem{Farina2019} D. Farina, V. Cavina, and V. Giovannetti, {\it Quantum Bath Statistical Tagging}, Phys. Rev. Lett. 100, 042327 (2019).
\bibitem{Jorgensen2020} M.R. J\o rgensen, P.P. Potts, M.G.A. Paris, J.B. Brask, {\it Tight bound on finite-resolution quantum thermometry at low temperatures}, arXiv:2001.04096 (2020).
\bibitem{Monras2011} A. Monras and F. Illuminati, {\it Measurement of damping and temperature: Precision bounds in Gaussian dissipative channels}, Phys. Rev. A 83, 012315 (2011).
\bibitem{Pinel2013} O. Pinel, P. Jian, N. Treps, C. Fabre, and D. Braun, {\it Quantum parameter estimation using general single-mode Gaussian states}, Phys. Rev. A 88, 040102 (2013).
\bibitem{Gao2014} Y. Gao, and H. Lee, {\it Bounds on quantum multiple-parameter estimation with Gaussian state}, Eur. Phys. J. D 68, 347 (2014).
\bibitem{Feyles2019} M.M. Feyles, L. Mancino, M. Sbroscia, I. Gianani, and M. Barbieri, {\it Dynamical role of quantum signatures in quantum thermometry}, Phys. Rev. A 99, 062114 (2019).
\bibitem{Ferraro2005} A. Ferraro, S. Olivares, M.G.A. Paris, {\it Gaussian states in continuous variable quantum information}, Bibliopolis, Napoli (2005).
\bibitem{Serafini2017} A. Serafini, {\it Quantum Continuous Variables: A primer of Theoretical Methods}, Taylor \& Francis, Oxford (2017).
\bibitem{Nichols2018} R. Nichols, P. Liuzzo-Scorpo, P.A. Knott, and G. Adesso, {\it Multiparameter Gaussian quantum metrology}, Phys. Rev. A 98, 012114 (2018). 
\bibitem{Hammerer} K. Hammerer, A. S. S{\o}rensen, and E. Polzik, {\it Quantum interface between light and atomic ensembles}, Rev. Mod. Phys. {\bf 82}, 1041 (2010). 
\bibitem{Genoni2016} M.G. Genoni, L. Lami, and A. Serafini, {\it Conditional and unconditional Gaussian quantum dynamics}, Contemporary Physics, 57:3, 331-249 (2016).
\bibitem{Breuer2002} H.-P. Breuer and F. Petruccione, {\it The Theory of Open Quantum Systems}, Oxford University Press on Demand, New York (2002).
\bibitem{Holstein1940} T. Holstein and H. Primakoff, {\it Field Dependence of the Intrinsic Domain Magnetization of a Ferromagnet}, Phys. Rev. 58, 1098 (1940).
\bibitem{Marcuzzi2016} M. Marcuzzi, J. Marino, A. Gambassi, and A. Silva, {\it Prethermalization from a low-density Holstein-Primakoff expansion}, Phys. Rev. B 94, 214304 (2016).
\bibitem{Paternostro2006} M. Paternostro, S. Gigan, M.S. Kim, F. Blaser, H.R. B\"ohm, and M. Aspelmeyer, {\it Reconstructing the dynamics of a movable mirror in a detuned optical cavity}, New J. Phys. 8, 107 (2006).
\bibitem{Hurwitz1964} A. Hurwitz, {\it Selected Papers on Mathematical Trends in Control Theory}, Ed. R. Bellman and R. Kalaba, New York: Dover (1964).
\bibitem{Bengtsson2006} I. Bengtsson and K. Zyczkowski, {\it Geometry of Quantum States: An Introduction to Quantum Entanglement}, Cambridge University Press, Cambridge (2006).
\bibitem{Pires2016} D.P. Pires, M. Cianciaruso, L.C. C\'{e}leri, G. Adesso, and D.O. Soares-Pinto, {\it Generalized Geometric Quantum Speed Limits}, Phys. Rev. X 6, 021031 (2016).
\bibitem{Morozova} E.A. Morozova, and N.N. Cÿencov, Markov Invariant Ge- ometry on Manifolds of states, J. Sov. Math. 56, 2648 (1991).
\bibitem{Petz1996a} D. Petz, and H. Hasegawa, On the Riemannian Metric of $\alpha$-Entropies of Density Matrices, Lett. Math. Phys. 38, 221 (1996).
\bibitem{Petz1996b} D. Petz, Monotone Metrics on Matrix Spaces, Linear Algebra Appl. 244, 81 (1996).
\bibitem{Petz2002} D. Petz, {\it Covariance and Fisher information in quantum mechanics}, J. Phys. A: Math. Gen. 35, 929 (2002).
\bibitem{uhlman1} A. Uhlmann, {\it Density operators as an arena for differential geometry},Rep. Math. Phys. {\bf 33}, 253 (1993).
\bibitem{uhlman2} A. Uhlmann, {\it Geometric phases and related structures}, Rep. Math. Phys. {\bf 36}, 461 (1995).
\bibitem{Monras2010} A. Monras and F. Illuminati, {\it Information geometry of Gaussian channels}, Phys. Rev. A 81, 062326 (2010).
\bibitem{Siudzinska2019} K. Siudzi\'{n}ska, K. Luoma, and W. T. Strunz, {\it Geometry on the manifold of Gaussian quantum channels}, Phys. Rev. A 100, 062308 (2019).
\bibitem{Petz1996} D. Petz and C. Sud\'{a}r, {\it Geometries of quantum states}, J. Math. Phys. 37, 2662 (1996).
\bibitem{Banchi2015} L. Banchi, S.L. Braunstein, and S. Pirandola, {\it Quantum fidelity for arbitrary Gaussian states}, Phys. Rev. Lett. 115, 260501 (2015).
\bibitem{Giedke2001} G. Giedke, B. Kraus, M. Lewenstein, and J. I. Cirac, {\it Entanglement Criteria for All Bipartite Gaussian States}, Phys. Rev. Lett. 87, 167904 (2001).

\bibitem{Giovannetti2004} V. Giovannetti, S. Lloyd, and L. Maccone, {\it Quantum-Enhanced Measurements: Beating the Standard Quantum Limit}, Science, 306, 5700, 1330-1336 (2004).
\bibitem{Giovannetti2006} V. Giovannetti, S. Lloyd, and L. Maccone, {\it Quantum Metrology}, Phys. Rev. Lett. 96, 010401 (2006).
 \end{thebibliography}
\end{document}